\documentclass[twocolumn,showpacs,preprintnumbers,amsmath,amssymb]{revtex4}
\usepackage{graphicx}
\usepackage{dcolumn}
\usepackage{bm}
\newcommand{\n}{\nonumber}

\newcommand{\dk}{{\rm d}{\bf k}}
\newcommand{\dx}{{\rm d}{\bf x}}

\newcommand{\vp}{\varphi}

\begin{document}

\title{Scale dependence of cosmological backreaction}

\author{Nan Li
\footnote{Electronic address: li at physik dot uni-bielefeld dot
de}}
\author{Dominik J. Schwarz
\footnote{Electronic address: dschwarz at physik dot uni-bielefeld
dot de}}

\affiliation{Fakult\"at f\"ur Physik, Universit\"at Bielefeld,
Universit\"atsstra\ss e 25, D-33615 Bielefeld, Germany}

\begin{abstract}
Due to the non-commutation of spatial averaging and temporal
evolution, inhomogeneities and anisotropies (cosmic structures)
influence the evolution of the averaged Universe via the
cosmological backreaction mechanism. We study the backreaction
effect as a function of averaging scale in a perturbative approach
up to higher orders. We calculate the hierarchy of the critical
scales, at which $10\%$ effects show up from averaging at different
orders. The dominant contribution comes from the averaged spatial
curvature, observable up to scales of $\sim 200~\mbox{Mpc}$. The
cosmic variance of the local Hubble rate is $10\%$ $(5\%)$ for
spherical regions of radius 40 $(60)~\mbox{Mpc}$. We compare our
result to the one from Newtonian cosmology and Hubble Space
Telescope Key Project data.
\end{abstract}

\pacs{95.36.+x, 98.65.Dx, 98.80.-k, 98.80.Jk}

\maketitle

Various cosmological observations, interpreted in the framework of
spatially flat, homogeneous and isotropic cosmogonies, have now
confirmed the accelerated expansion of the Universe. The most direct
evidence comes from the study of supernova (SN) of type
Ia~\cite{Astier:2005qq}. Many attempts have been proposed to
understand this mystery, e.g., dark energy in the form of a
cosmological constant, quintessence field or modification of
gravity. However, these suggestions always rely on the homogeneity
and isotropy of the cosmic medium, which are rather rough
approximations.

The Universe hosts enormous structures. In our neighborhood, there
seem to exist two voids, both $35$ to $70~\mbox{Mpc}$ across,
associated with the so-called velocity anomaly~\cite{Rizzi:2007th},
a large filament known as the Sloan great wall about
$400~\mbox{Mpc}$ long~\cite{Gott:2003pf} and the Shapely
supercluster with a core diameter of $40~\mbox{Mpc}$ at a distance
of $\sim 200~\mbox{Mpc}$ from us~\cite{SSC}. Furthermore, based on
the Hubble Space Telescope (HST) Key Project
data~\cite{Freedman:2000cf}, evidence for a significant anisotropy
in the local Hubble expansion at distances of $\sim 100~\mbox{Mpc}$
was found~\cite{McClure:2007vv}, and an anisotropy of SN Ia Hubble
diagrams extending to larger distances has been reported
recently~\cite{Schwarz:2007wf}. Therefore, spatial homogeneity and
isotropy seem to be valid only on scales larger than $\sim
100~\mbox{Mpc}$~\cite{Hogg:2004vw}, and effects of local
inhomogeneities are worthy of investigation. More specifically,
observables from within a few $100~\mbox{Mpc}$ must be revisited
critically. The most fundamental of those are cosmic distances and
the Hubble constant $H_0$.

In this paper, we study the averaging of the inhomogeneous and
anisotropic Universe over a local domain in space-time. We stick to
the idea of cosmological inflation, assuming that the Universe
approaches homogeneity and isotropy at scales as large as the Hubble
distance.

Many cosmological observables are averaged quantities. For instance,
the matter power spectrum is a Fourier transform and thus a volume
average weighted by a factor $e^{i{\bf k}\cdot{\bf x}}$. Another
very important example is the idealized measurement of
$H_0$~\cite{Tully:2007ue}. One picks $N$ standard candles in a local
volume $V$ (e.g., SN Ia in the Milky Way's neighborhood out to
$\sim100~\mbox{Mpc}$), measures their luminosity distances $d_i$ and
recession velocities $v_i=cz_i$ ($z_i$ being the redshift of each
candle) and performs the average $H_0 \equiv \frac{1}{N}\sum_{i=1}^N
\frac{v_i}{d_i}$. In the limit of a very big sample, it turns into a
volume average $H_0=\frac{1}{V}\int\frac{v}{d}{\rm d}V$.

Cosmological observations are made on the past light-cone, so one
should average over a light-cone volume. However, for objects at $z
\ll 1$, spatial averaging on a constant-time-hypersurface is a good
approximation, as the Universe does not change significantly on the
temporal scale involved.

Due to the nonlinearity of the Einstein equations, spatial averaging
and temporal evolution do not commute. Hence, inhomogeneities and
anisotropies affect the evolution of the averaged Universe via the
so-called ``backreaction
mechanism"~\cite{Ellis,Russ,Buchert:1999er,Zimdahl:2000zm,
Buchert:2006,Mattsson:2007qp,Li:2007ci}. Below, we utilize Buchert's
averaging method~\cite{Buchert:1999er} to estimate the order of
magnitude of backreaction effects and study the signatures of
averaging from the local measurement of $H_0$.

Buchert's setup is well adapted to the situation of a real observer,
if we are allowed to neglect the difference between baryons and cold
dark matter (CDM). On scales $\gtrsim 10~\mbox{Mpc}$, baryon
pressure is insignificant, and a real observer comoves with matter,
uses her own clock and regards space to be time-orthogonal. These
conditions define a comoving synchronous coordinate system. There
are no primordial vector perturbations from cosmological inflation,
so we assume the Universe to be irrotational. As we are concerned
about the present Universe, radiation is thus neglected. Moreover,
the cosmological constant is also supposed to vanish, as we ask
whether averaging could mimic a component of dark energy. Following
Buchert, we use physically comoving boundaries to thoroughly fix the
averaging procedure.

In the synchronous coordinates, the metric of the inhomogeneous and
anisotropic Universe is $\mbox{d}s^2 = -\mbox{d}t^2 + g_{ij}(t,{\bf
x})\mbox{d}x^i\mbox{d}x^j$. The spatial average of an observable
$O(t,\bf x)$ at time $t$ is defined as
\begin{eqnarray}
\langle O \rangle_D\equiv \frac{1}{V_D(t)}\int_D O(t,{\bf x})
\sqrt{\mbox{det}g_{ij}}\mbox{d}\bf x . \label{average}
\end{eqnarray}
$V_D(t) \equiv \int_D \sqrt{\mbox{det}g_{ij}} \mbox{d}{\bf x}$ is
the volume of a comoving domain $D$, introducing an effective scale
factor
\begin{eqnarray}
\frac{a_D}{a_{D_0}}\equiv
\left(\frac{V_D}{V_{D_0}}\right)^{1/3}. \label{ad}
\end{eqnarray}
The subscript 0 denotes the present time. The effective Hubble rate
is thus defined as $H_D\equiv \dot{a}_D/a_D=\langle
\theta\rangle_D/3$ ($\theta$ being the volume expansion
rate)~\cite{Buchert:1999er}.

Effective Friedmann equations for a dust Universe follow from
averaging Einstein's equations~\cite{Buchert:1999er},
\begin{eqnarray}
\left(\frac{\dot{a}_D}{a_D}\right)^2= \frac{8\pi G}{3}\rho_{\rm
eff}, \quad -\frac{\ddot{a}_D}{a_D} = \frac{4\pi G}{3}(\rho_{\rm
eff}+3p_{\rm eff}). \label{buchert}
\end{eqnarray}
Here $\rho_{\rm eff}$ and $p_{\rm eff}$ are the energy density and
pressure of an effective fluid,
\begin{eqnarray}
&&\rho_{\rm eff} \equiv \langle \rho\rangle_D-\frac{1}{16\pi G}
\left(\langle Q\rangle_D+\langle {\cal R}\rangle_D\right), \label{rhoeff} \\
&&p_{\rm eff} \equiv - \frac{1}{16\pi G}\left(\langle Q\rangle_D-
\frac{1}{3}\langle {\cal R}\rangle_D\right), \label{peff}
\end{eqnarray}
where $\rho$ is the energy density of dust. $\langle
Q\rangle_D\equiv\frac{2}{3}(\langle \theta^2\rangle_D-\langle
\theta\rangle_D^2)-2\langle \sigma^2\rangle_D$ denotes the
kinematical backreaction ($\sigma^2$ being the shear scalar) and
$\langle {\cal R}\rangle_D$ the averaged spatial curvature. They are
related by an integrability condition~\cite{Buchert:1999er},
\begin{eqnarray}
(a_D^6\langle Q\rangle_D)^{^{\textbf{.}}}+a_D^4(a_D^2\langle {\cal
R}\rangle_D)^{^{\textbf{.}}}=0.\label{int}
\end{eqnarray}
We further define an effective equation of state,
\begin{equation}
w_{\rm eff}\equiv \frac{p_{\rm eff}}{\rho_{\rm eff}}=\frac{\langle
{\cal R}\rangle_D-3\langle Q\rangle_D}{2\langle
\theta\rangle_D^2}.\n
\end{equation}
So we find that cosmological backreaction gives rise to a nontrivial
equation of state, even for a dust Universe~\cite{Li:2007ci}.

Alternatively, we may map this effective fluid on a model with dust
and dark energy. Let $n$ be the number density of dust particles,
and $m$ be their mass. For any comoving domain, $\langle
n\rangle_D=\langle n\rangle_{D_0}(a_{D_0}/a_D)^3$. In the dust
Universe, $\rho(t, {\bf x}) \equiv mn(t, {\bf x})$, and we identify
$\rho_{\rm m} \equiv \langle \rho\rangle_{D}= m \langle
n\rangle_{D}$. From Eq.~(\ref{rhoeff}), dark energy is consequently
$\rho_{\mathrm{de}}=-(\langle Q\rangle_D+\langle {\cal
R}\rangle_D)/(16\pi G)$, with the relevant equation of state reading
\begin{eqnarray}
w_{\mathrm{de}}\equiv\frac{p_{\mathrm{de}}}{\rho_{\mathrm{de}}}
=\frac{p_{\mathrm{eff}}}{\rho_{\mathrm{de}}}=-\frac{1}{3}+\frac{4\langle
Q\rangle_D}{3(\langle Q\rangle_D+\langle {\cal R}\rangle_D)}.\n
\end{eqnarray}
It is $-1$, iff $\langle Q\rangle_D=-\frac{1}{3}\langle
R\rangle_D$~\cite{Buchert:2006}, corresponding to a cosmological
constant $\Lambda=\langle Q\rangle_D$.

Equations (\ref{buchert}) and (\ref{int}) are not closed, as the
four unknown variables $\langle Q\rangle_D$, $\langle{\cal
R}\rangle_D$, $\langle \rho\rangle_D$ and $a_D$ are constrained by
only three equations. Below, we close these dynamical equations for
the averaged Universe by means of cosmological perturbation theory.

We wish to estimate the  scale dependence of $\langle Q\rangle_D$,
$\langle {\cal R}\rangle_D$, $\langle \rho\rangle_D$, $H_D$ and
$w_{\mathrm{eff}}$. We start from a spatially flat dust model. In
the comoving synchronous gauge, the linear perturbed metric is
$\mbox{d}s^2=-\mbox{d}t^2+a^2(t)[(1-2\Psi)\delta_{ij}+
(\partial_{i}\partial_{j}-\frac{1}{3}\delta_{ij}\Delta)\chi]\mbox{d}x^i
\mbox{d}x^j$. Here, its scale factor $a(t)$ ($a_0\equiv 1$) is
different from $a_D$, and their relation was provided in
Ref.~\cite{Li:2007ci}. $\Psi$ and $\chi$ are the scalar metric
perturbations, and $\Delta$ is the three-dimensional Laplace
operator. The solutions for $\Psi$ and $\chi$ are given in terms of
the time-independent peculiar gravitational potential $\varphi({\bf
x})$: $\Psi=\frac{1}{2}\Delta \varphi
t_0^{4/3}t^{2/3}+\frac{5}{3}\varphi$ and $\chi=-3\varphi
t_0^{4/3}t^{2/3}$ (only growing modes are taken into
account)~\cite{Li:2007ci}. Moreover, $\varphi$ is related to the
hypersurface-invariant variable $\zeta$~\cite{Bardeen:1988hy} by
$\zeta=\frac{1}{2}\Delta \varphi
t_0^{4/3}t^{2/3}-\frac{5}{3}\varphi$.

Following Ref.~\cite{Li:2007ci}, we use the metric perturbations
attained from linear perturbation theory together with the
non-perturbative integrability condition to obtain the averaged
physical observables up to second order. We focus on the dominant
contributions from the growing modes and neglect the decaying ones,
since we are interested in the late time effects of cosmic
averaging. Thus, we find
\begin{eqnarray}
\langle Q\rangle_D &=& \frac{a_{D_0}}{a_D}B(\varphi)t^2_0,\label{q2}\\
\langle {\cal R}\rangle_D &=&
\frac{20}{3}\frac{a_{D_0}^2}{a_D^2}\langle
\Delta \varphi\rangle-5\frac{a_{D_0}}{a_D}B(\varphi)t^2_0,\label{r2}\\
\langle \rho\rangle_D&=&\frac{1}{6\pi G t_0^2}\frac{a_{D_0}^3}{a_D^3},\label{rho2}\\
H_D&=&\frac{2}{3t_0}\frac{a_{D_0}^{3/2}}{a_D^{3/2}}
\left[1-\frac{5}{4}\frac{a_D}{a_{D_0}}t_0^2\langle \Delta \varphi\rangle\right.\n\\
&&\left. +\frac{3}{4}\frac{a_D^2}{a_{D_0}^2}t_0^4\left(B(\varphi)
-\frac{25}{24}\langle\Delta\varphi\rangle^2\right)\right],\label{theta2}\\
w_{\rm eff}&=&\frac{5}{6}\frac{a_D}{a_{D_0}}t_0^2\langle\Delta
\varphi\rangle-\frac{a_D^2}{a_{D_0}^2}t_0^4\left(B(\varphi)-\frac{25}{12}\langle
\Delta \varphi\rangle^2\right),\n\\\label{w2}
\end{eqnarray}
with
$B(\varphi)\equiv\langle\partial^i(\partial_i\varphi\Delta\varphi)
-\partial^i(\partial_j\varphi\partial^j\partial_i\varphi)\rangle
-\frac{2}{3}\langle \Delta \varphi\rangle^2$ and $\langle
O\rangle\equiv\int_D O {\rm d}{\bf x}/ \int_D {\rm d}{\bf x}$. We
see from Eqs.~(\ref{q2}) -- (\ref{w2}) that these quantities are
polynomials of surface terms. Thus, all information is encoded on
the boundaries of the comoving domain $D$. The temporal dependence
of these averaged quantities can be found in Ref.~\cite{Li:2007ci},
and their leading terms are gauge-invariant~\cite{Li:2007ci}.

Our perturbative results suggest to write $\langle Q\rangle_D$ and
$\langle {\cal R}\rangle_D$ in a Laurent series of $a_D$. (Recently,
a power-law ansatz for the integrability condition was investigated
in Ref.~\cite{Buchert:2006ya}.) We know from Eqs.~(\ref{q2}) and
(\ref{r2}) that $\langle Q\rangle_D$ and $\langle {\cal R}\rangle_D$
start from different powers: $a_D^{-1}$ and $a_D^{-2}$, so $\langle
Q\rangle_D=\sum_{n=-1}Q_n(\frac{a_D}{a_{D_0}})^n$ and $\langle {\cal
R}\rangle_D=\sum_{n=-2}{\cal R}_n(\frac{a_D}{a_{D_0}})^n$. The
integrability condition then connects the coefficients:
$(n+6)Q_n+(n+2){\cal R}_n=0$. Thus, $Q_0=-\frac{1}{3}{\cal R}_0$ at
third order in perturbation theory. Therefore, cosmological
backreaction can mimic a cosmological constant, but induces extra
terms as well. The third order results will be presented
elsewhere~\cite{third}.

The effect of cosmological backreaction in the early Universe is
tiny and is undistinguishable from that of a homogeneous curvature,
as $w_{\mathrm{de}}\rightarrow -1/3$ when $a_D\rightarrow 0$. This
result seems inconsistent with our intuition of a vanishing
cosmological backreaction at early times, suggesting that $w_{\rm
de}$ should also vanish. However, as we have seen above,
cosmological averaging gives rise to extra degrees of freedom in the
dynamics of the averaged Universe.

The effect of averaging over a typical domain is provided by the
ensemble average. From Eq.~(\ref{w2}), we find $\overline{w_{\rm
eff}}=\frac{11}{4}(\frac{a_D}{a_{D_0}})^2t_0^4\overline{\langle
\Delta \varphi\rangle^2}>0$. This means that cosmological
backreaction is expected to lead to a positive definite equation of
state. However, we should pay attention that $\overline{w_{\rm
eff}}$ is of second order, but the root of its variance $[{\rm
Var}(w_{\rm eff})]^{1/2}=\frac{5}{6}\frac{a_D}{a_{D_0}}t_0^2
(\overline{\langle\Delta\varphi\rangle^2})^{1/2}$ is of first order
and therefore larger than $\overline{w_{\rm eff}}$. Thus, the
possibilities of $w_{\rm eff} < 0$ and the effective acceleration of
the averaged Universe cannot be easily excluded. Looking only at
mean values of the ensemble obviously causes an underestimation of
the possible backreaction effects, as we often observe just one
particular domain in the Universe.

We now turn to estimate the order of magnitude of cosmological
backreaction as a function of the averaging scale $r\sim
V_{D_{0}}^{1/3}$. We show that cosmological averaging produces
important modifications to local physical observables and determine
the averaging scale, at which corrections show up at a $10\%$ level.

Effective acceleration of the averaged Universe occurs if
$\rho_{\mathrm{eff}}+3p_{\mathrm{eff}}<0$, i.e., $\langle
Q\rangle_D>4\pi G\langle \rho\rangle_D$. From Eqs.~(\ref{q2}) and
(\ref{rho2}), we have
\begin{eqnarray}
\left|\frac{\langle Q\rangle_D}{4\pi G\langle \rho\rangle_D}\right|=
\frac{3}{2}\frac{a_D^2}{a_{D_0}^2}B(\varphi)t_0^4
=\frac{8}{27}\frac{R_{\rm H}^4}{(1+z)^2}B(\varphi),\label{estimate}
\end{eqnarray}
with $R_{\rm H}=2.998\times 10^3h^{-1}~\mbox{Mpc}$ being the present
Hubble distance. In Eq.~(\ref{estimate}), we can safely use the
results for the background Unverse: $a_D/a_{D_0}=1/(1+z)$ and
$t_0=2R_{\rm H}/3$, because $B(\varphi)$ is of second order. Since
the ratio in Eq.~(\ref{estimate}) is dimensionless, a dimensional
analysis immediately implies $|\langle Q\rangle_D/4\pi G\langle
\rho\rangle_D|\propto(R_{\rm H}/r)^4$, where for an almost
scale-invariant power spectrum the unique relevant
scale is the averaging scale $r$. The order of magnitude of
Eq.~(\ref{estimate}) can be estimated as
\begin{eqnarray}
\left|\frac{\langle Q\rangle_D}{4\pi G\langle
\rho\rangle_D}\right|\sim\frac{8}{75}\frac{1}{(1+z)^2}\left(\frac{R_{\rm
H}}{r}\right)^4 \mathcal{P}_{\zeta}.\label{onsetq}
\end{eqnarray}
$\mathcal{P}_{\zeta}=2.457\times 10^{-9}$ is the dimensionless power
spectrum~\cite{Komatsu}. We pick the second term in $B(\vp)$,
$\langle\partial^i(\partial_j\varphi\partial^j\partial_i\varphi)\rangle$,
to demonstrate how to obtain this estimate. In the Fourier space,
$\partial^i\varphi\rightarrow ik^i\varphi\sim\varphi/r$. The latter
step comes from the observation that only structure of the size of
the averaged volume cannot be averaged out. At much smaller scales,
structures contribute a negligible amount to $B(\varphi)$, because
it is not positive definite and is expected to fluctuate on small
scales. Thus,
\begin{eqnarray}
\langle\partial^i(\partial_j\varphi\partial^j\partial_i\varphi)\rangle\rightarrow
\frac{1}{r^4}\langle\varphi^2\rangle\sim\frac{1}{r^4}\mathcal{P}_{\varphi}=
\frac{9}{25}\frac{1}{r^4}\mathcal{P}_{\zeta},\nonumber
\end{eqnarray}
i.e., each derivative in $B(\varphi)$ contributes a factor $1/r$. Also,
$\langle\varphi^2\rangle$ in the Fourier space is estimated as the
power spectrum $\mathcal{P}_{\varphi}$. Since $\varphi$ is constant
in time, and $\zeta\approx-5\varphi/3$ on superhorizon scales, we
can identify today's $\mathcal{P}_{\varphi}$ with
$9\mathcal{P}_{\zeta}/25$. Similar estimation works for the other
two terms in $B(\varphi)$.

The kinematical backreaction induces $10\%$ and larger modifications
if $|\langle Q\rangle_D/ 4\pi G\langle \rho\rangle_D|\gtrsim 0.1$.
This happens if
\begin{eqnarray}
r_Q\lesssim\frac{21 h^{-1}}{\sqrt{1+z}}~\mbox{Mpc}. \label{rq}
\end{eqnarray}
For observations at $z\ll 1$,  $r_Q\lesssim 30\, \mbox{Mpc}$ ($h =
0.7$).

The averaged spatial curvature $\langle {\cal R}\rangle_D$ is the
most important correction to energy density. The criterion for the
scale, at which its effect emerges, is estimated analogously by
\begin{equation}
\left|\frac{\rho_{\rm eff}}{\langle \rho \rangle_D}-1\right| \approx
\left|\frac{\langle{\cal R}\rangle_D}{16\pi G\langle
\rho\rangle_D}\right| \sim
\frac{2}{3}\frac{1}{1+z}\left(\frac{R_{\rm H}}{r}\!\right)^2
\sqrt{\mathcal{P}_{\zeta}}.\label{onsetr}
\end{equation}
We find effects larger than $10\%$ within
\begin{eqnarray}
r_{\cal R}\lesssim\frac{54h^{-1}}{\sqrt{1+z}}~\mbox{Mpc}. \label{rr}
\end{eqnarray}
At small redshifts, $r_{\cal R}\lesssim 77~\mbox{Mpc}$. Furthermore,
effects above $1\%$ are expected up to a scale of $\sim
240~\mbox{Mpc}$. Note that the curvature of the Universe has been
measured at the few per cent accuracy in the cosmic microwave
background (CMB)~\cite{Komatsu}. It was shown in
Ref.~\cite{Clarkson} that even small curvature might affect the
analysis of high-$z$ SNe significantly.

Finally, we turn to the Hubble rate. To go beyond the order of
magnitude estimates above, we calculate the ensemble mean and its
variance (cosmic variance) of the relative fluctuation of the Hubble
rate $\delta_H \equiv (H_D-H_0)/H_0$. Before doing so, let us stress
that the analogous order of magnitude estimate for $\delta_H$
agrees with the result for $[{\rm Var}(\delta_H)]^{1/2}$ given below
up to a factor of $\sim 2$. For a spherical domain of
radius $r$, we find from Eq.~(\ref{theta2}),
\begin{equation}
\overline{\delta_H}=-\frac{41}{32}\frac{a_D^2}{a_{D_0}^2}t_0^4
\overline{\langle\Delta\varphi\rangle^2}, \quad {\rm
Var}\left(\delta_H\right)=\frac{25}{16}\frac{a_D^2}{a_{D_0}^2}t_0^4
\overline{\langle\Delta\varphi\rangle^2}, \label{deltaH}
\end{equation}
where
\begin{eqnarray}
\overline{\langle\Delta\varphi\rangle^2}&=&\int\frac{\dx_1\dx_2}{V^2}\frac{\dk_1\dk_2}{(2\pi)^{6}}
k_1^2k_2^2\overline{\varphi_{{\bf k}_1}\varphi_{{\bf k}_2}}e^{i({\bf
k}_1\cdot{\bf x}_1+{\bf k}_2\cdot{\bf x}_2)} \n\\
&=&\int\frac{\dx_1\dx_2}{V^2}\frac{\dk}{32\pi^4}k{\cal
P}_{\varphi}(k)e^{i{\bf k}\cdot({\bf x}_1+{\bf x}_2)},\n
\end{eqnarray}
with $V=4\pi r^3/3$ (a top-hat window). Above, we introduce the
dimensionless power spectrum as $\overline{\varphi_{{\bf
k}_1}\varphi_{{\bf k}_2}}\equiv2\pi^2\delta({\bf k}_1+{\bf
k}_2){\cal P}_{\varphi}(k_1)/k_1^3$ ($k\equiv |{\bf k}|$). So
\begin{equation}
{\rm
Var}\left(\delta_H\right)=\frac{25}{144\pi^2}\frac{1}{(1+z)^2}\left(\frac{R_{\rm
H}}{r}\right)^4\int_0^{\infty}{\rm d}x{\cal
P}_{\varphi}(x/r)J^2_{3/2}(x).\label{variance}
\end{equation}
$J_{3/2}(x)$ is the Bessel function of first kind ($x\equiv kr$).
For a scale-invariant power spectrum, we must introduce an
ultraviolet cutoff ${\cal P}_{\varphi}(k)={\cal
P}_{\varphi}e^{-k/k_{\rm c}}$. No cutoff is required for a
red-tilted spectrum ${\cal P}_{\varphi}(k)={\cal
P}_{\varphi}(k/k_0)^{n_{\rm s}-1}$ ($n_{\rm s}<1$ being the spectrum
index), consistent with WMAP5~\cite{Komatsu}. Here, let us stress
that although $[{\rm Var}\left(\delta_H \right)]^{1/2}$ is only a
first order quantity, the next contribution is already of third
order, if we consult the perturbed metric to second order. Since we
constrain our attention to the leading order effects, these higher
order terms are negligible~\cite{third}.

Now we can link the effect of cosmological backreaction in Buchert's
setup (evaluated in a perturbative approach up to second order) to
actual cosmological observations. The trick is to consider the scale
dependence but not the time dependence. The value of the relative
fluctuation of the Hubble rate in Eq.~(\ref{deltaH}) is dominated by
its variance, and thus the sign of the observed value of $\delta_H$
cannot be predicted. A comparison of the mean and the root of the
variance of $\delta_H$ tells us that perturbation theory breaks down
below $\sim20$~Mpc.

The scale dependence of the cosmic variance of $\delta_H$ has
previously been studied in the context of Newtonian
cosmology~\cite{Turner,Shi}, largely based on CDM simulations. In
this setting, the variance of $\delta_H$ is due to peculiar motions
(besides sampling variance and observational errors). In a
relativistic and comoving approach, peculiar velocities vanish
identically, and the cosmic variance of the Hubble rate turns into a
curvature effect, because Eqs.~(\ref{r2}) and (\ref{deltaH}) give
${\rm Var}(\delta_H) \propto\overline{\langle {\cal R}\rangle_D^2}$.

\begin{figure}
\centerline{
\includegraphics[width=0.7\linewidth,angle=270]{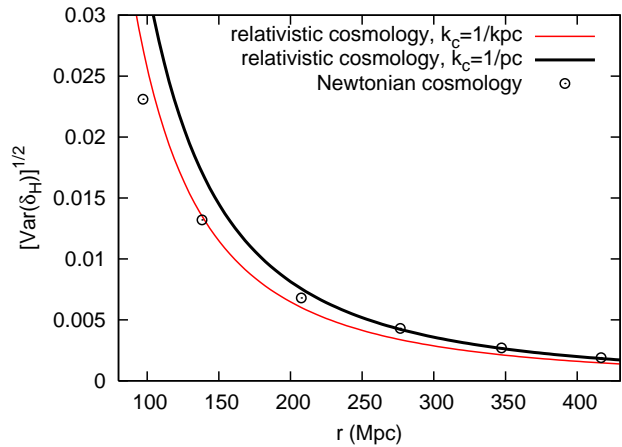}}
\caption{Scale dependence of the cosmic variance of the Hubble rate.
Data are from the Newtonian CDM model in Ref.~\cite{Shi}, with
$h=0.5$, $\Omega_{\rm m}=1$ and a COBE-normalized power spectrum.
Thick and thin lines correspond to the relativistic result
Eq.~(\ref{variance}) for a scale-invariant power spectrum with
cutoffs at $k_{\rm c}$=1/kpc (simulation) and 1/pc (physical),
respectively.}\label{fig1}
\end{figure}

In Fig.~(\ref{fig1}), we compare the relativistic (correct up to
second order) result Eq.~(\ref{variance}) to Newtonian ``standard
CDM" case in Ref.~\cite{Shi}. We find that up to $\sim400~{\rm
Mpc}$, our results for scale-invariant power spectra ($k_{\rm c}=
1$/kpc corresponding to a typical cutoff in CDM simulations and 1/pc
to the physical cutoff in the primordial CDM spectrum) agree with
Newtonian simulations. This agreement is not unexpected, as metric
perturbations and peculiar velocities are small at
$\sim100~\mbox{Mpc}$ scales.

The consistency between the relativistic and Newtonian approaches
encourages the comparison of our perturbative results with
experimental data. We compare Eq.~(\ref{variance}) with observations
from the HST Key Project~\cite{Freedman:2000cf}. We use 54
individual measurements of $H_0$ in the CMB rest frame (corrected
for local flow) from SN Ia and the Tully-Fisher relation (Tabs.~(6)
and (7) in Ref.~\cite{Freedman:2000cf}). We have checked explicitly
that the SN and Tully-Fisher measurements of $H_0$ are consistent
with each other, while we cannot confirm that for the fundamental
plane method and thus dropped them from a former analysis.

We restrict our analysis to objects between $31.3$ to
$467.0~\mbox{Mpc}$, as Eq.~(\ref{variance}) can be trusted only
above $30~\mbox{Mpc}$. Be $r_i$, $H_i$ and $\sigma_i$ the distance,
Hubble rate and $1\sigma$ error for the $i'$th datum, with distances
increasing. We calculate the mean distance for the nearest $k$
objects by $\bar{r}_k=\sum_{i=1}^k g_i r_i/\sum_{i=1}^k g_i$, with
weights $g_i=H_0^2/\sigma_i^2$. An analogue holds for the averaged
Hubble rate $\bar{H}_k$, i.e., $H_D$ for different subsets. The
empirical variance of each subset is $\bar{\sigma}^2_k=[\sum_{i=1}^k
g_i (H_i-\bar{H}_k)^2]/[H_0^2(k-1)\sum_{i=1}^k g_i]$. Notice that
Eq.~(\ref{variance}) is insensitive to global calibration issues.

\begin{figure}
\centerline{
\includegraphics[width=0.7\linewidth,angle=270]{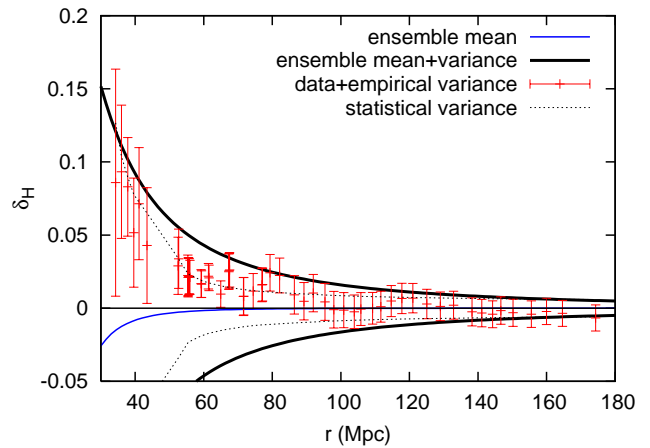}}
\caption{Relative fluctuation of the Hubble rate from cosmological
backreaction and its cosmic variance band (thick lines) compared to
the empirical mean and variance of $\delta_H$ obtained from the HST
Key Project data~\cite{Freedman:2000cf} as a function of averaging
radius. The thin line shows the ensemble mean of $\delta_H$. The
band enclosed by the thick lines indicates the effect of the
inhomogeneities ($\propto 1/r^2$), and the dashed lines are the
effect from sampling with given measurement errors in an otherwise
perfectly homogeneous Universe.} \label{fig2}
\end{figure}

The comparison of the result Eq.~(\ref{variance}) with the HST Key
Project data is shown in Fig.~(\ref{fig2}). We now normalize to the
WMAP5 best-fit power-law spectrum, with pivot $k_0$=0.002/Mpc and
spectral index $n_{\rm s}=0.960$ and use
$H_0=72$~km/s/Mpc~\cite{Komatsu}. We see that the theoretical band
matches the experimental data well, without any fit parameter in the
panel. Moreover, we see from Fig.~(\ref{fig2}) that the value of
$\delta_H$ is positive within $\sim100$~Mpc. This is consistent with
the result in a recent paper~\cite{Hunt} that we are located in a
200 -- 300~Mpc underdense void, which is expanding faster than the
global Hubble rate.

Before we can claim that we have observed the expected $1/r^2$
behavior in Eq.~(\ref{variance}) and thus the evidence for
cosmological backreaction, we must make sure that statistical noise
cannot account for it. In the case of a perfectly homogeneous
coverage of the averaged domain with standard candles, we would
expect a $1/r^{3/2}$ behavior. In Fig.~(\ref{fig2}), we show the
statistical noise for the actual data set ($1/(\sum_{i=1}^k
g_i)^{1/2}$), which is smaller than our result Eq.~(\ref{variance}).
It turns out that the sampling noise for this small data set is
still too large to claim that the inhomogeneity of the Universe can
be detected in the relative fluctuation of the Hubble rate observed
by the HST Key Project. However, it is fully consistent with our
theoretical expectations. Actually the fluctuation $\delta_H$
appears to be smaller than expected, and one might wonder why that
is so. From the theoretical expectation plotted in
Fig.~(\ref{fig2}), we find that at $\sim 40$ $(60)~\mbox{Mpc}$, the
value of $H_D$ differs from its global value $72~{\rm km/s/Mpc}$
(WMAP5) by about $10\%$ $(5\%)$, whereas the expected variance for a
perfectly homogeneous and isotropic Universe is $8\%$ $(2\%)$.

A similar analysis of the Hubble diagram was pioneered in
Refs.~\cite{Cooray:2006ft,Hui:2005nm,Hau,Durrer}, in which the
velocity field of the local Universe and its influence on the
correlated fluctuations in luminosity distance and the Hubble rate
was analyzed. Two essential differences to this work are that our
analysis includes effects to higher orders and we study the scale
dependence of the averaged observables. Although the relative
fluctuation of the Hubble rate was not explicitly analyzed in
Refs.~\cite{Cooray:2006ft,Hui:2005nm,Hau}, it seems to us that our
results are consistent with those findings.

To summarize, we argue that cosmological averaging (backreaction)
gives rise to observable effects up to scales of $\sim
200~\mbox{Mpc}$. However, it is not sufficient to explain the
observed accelerated expansion at this point.

We find a hierarchy of backreaction effects. The averaged spatial
curvature $\langle{\cal R}\rangle_D$ leads to $10\%$ $(1\%)$ effects
up to $\sim80$ $(240)~\mbox{Mpc}$ in a dust model with $h=0.7$.
Below $\sim40~\mbox{Mpc}$, the cosmic variance of the Hubble rate is
larger than $10\%$, which coincides with the estimate from the
effect of peculiar motions in Newtonian setup. Within
$\sim30~\mbox{Mpc}$, the kinematical backreaction $\langle
Q\rangle_D$, due to second order perturbations caused by local
inhomogeneities and anisotropies, enters the game. Cosmological
backreaction may put some of the steps on the cosmological distance
ladder in question, as they are deeply in the domain of large
backreaction, i.e., a large fluctuation between small averaged
volumes.

Our findings call for revisiting local observations, like galaxy
redshift surveys, in terms of possible backreaction signatures. The
large scale physics of primordial CMB anisotropies is not affected.
However, this statement cannot be made for secondary effects, e.g.,
the late integrated Sachs-Wolfe effect.

We are grateful to Thomas Buchert, Hengtong Ding, Stefan
Fr\"{o}hlich, Florian K\"{u}hnel, Julien Larena, Megan McClure,
Chuan Miao, Aseem Paranjape, Aleksandar Raki\'{c}, Marina Seikel,
Tejinder P. Singh, Glenn Starkman and David L. Wiltshire for
discussions. The work of N.L. is supported by the DFG under grant
GRK 881.

\end{document}